\documentclass[aps,twocolumn,superscriptaddress,amsfont,amssymb,amsmath,showpacs,balancelastpage,english]{revtex4-1} 
\usepackage{amsmath}
\usepackage{amssymb}
\usepackage{babel}
\usepackage{graphicx}
\usepackage{color}
\usepackage{natbib}
\usepackage{mathrsfs}
\usepackage{longtable}



\begin{document}

\title{Constraining the propagation speed of gravitational waves with compact binaries at cosmological distances}

\author{Atsushi Nishizawa}
\email{anishi@caltech.edu}
\affiliation{Theoretical Astrophysics 350-17, California Institute of Technology, Pasadena, California 91125, USA}

\begin{abstract}%%%%%%%%%%%%%%%%%%%%%%%%%%%%%%%%%%%%%%%
In testing gravity a model-independent way, one of crucial tests is measuring the propagation speed of a gravitational wave (GW). In general relativity, a GW propagates with the speed of light, while in the alternative theories of gravity the propagation speed could deviate from the speed of light due to the modification of gravity or spacetime structure at a quantum level. Previously we proposed the method measuring the GW speed by directly comparing the arrival times between a GW and a photon from the binary merger of neutron stars or neutron star and black hole, assuming that it is associated with a short gamma-ray burst. The sensitivity is limited by the intrinsic time delay between a GW and a photon at the source. In this paper, we extend the method to distinguish the intrinsic time delay from the true signal caused by anomalous GW speed with multiple events at cosmological distances, also considering the redshift distribution of GW sources, redshift-dependent GW propagation speed, and the statistics of intrinsic time delays. We show that an advanced GW detector such as Einstein Telescope will constrain the GW propagation speed at the precision of $\sim 10^{-16}$. We also discuss the optimal statistic to measure the GW speed, performing numerical simulations. 
\end{abstract}

\date{\today}

\maketitle

%%%%%%%%%%%%%%%%%%%%%%%%%%%%%%%%%%%%%%%%%%%%%

\section{Introduction}

The second-generation laser-interferometric gravitational wave (GW) detectors would accomplish the first detection of a GW in the coming a few years and open up GW astronomy \cite{Evans2014GRG}. After that, the detections of multiple events at cosmological distance would be realized with the third-generation ground-based GW detector such as Einstein telescope (ET) \cite{Punturo:2010zza} and 40-km LIGO \cite{Dwyer2015PRD}. The GW observations enable us not only to gain information about astronomical objects and cosmology \cite{Sathyaprakash:2009xs} but also to test gravity theories in strong and dynamical regimes of gravity (for reviews, see \cite{Will:2014kxa,Yunes:2013dva,Gair:2012nm,Berti2015CQG}). 

To test gravity with GWs, it is crucial to search for anomalous deviation from general relativity (GR) in a model-independent way. There have been many suggestions of such methods: seeking for the deviation from GR in GW phase evolution of compact-binary inspiraling \cite{Mishra:2010tp,Li:2011cg,Yunes:2009ke,Cornish:2011ys} and in GW waveforms of black-hole ringdown \cite{Gossan2012PRD,Meidam2014PRD}, and non-GR GW polarizations \cite{Seto:2007tn,Nishizawa:2009bf,Hayama:2012au,Chatziioannou:2012rf}. One of other tests is measuring the propagation speed of a GW. In GR, a GW propagates with the speed of light \footnote{Even in GR, the GW propagation speed could seemingly deviate from the speed of light due to the backscattering of gravitons by spacetime curvature, which is so-called the tail effect \cite{Malec2005CQG}. However, this effect is efficient only for GW whose wavelength is cosmological horizon scale in the matter-dominated era and would be irrelevant in direct detection experiments of GW.}, while in the alternative theories of gravity the propagation speed could deviate from the speed of light due to the modification of gravity (see \cite{Saltas:2014dha,Bellini2014JCAP,Gleyzes2014IJMPD,Ballesteros2015JCAP} for general formulations, and for more specific cases, nonzero graviton mass \cite{Gumrukcuoglu:2011zh,DeFelice:2013nba} and extra dimensions \cite{Sefiedgar:2010we}). Also the modification of spacetime structure at a quantum level may affect the propagation of a GW \cite{AmelinoCamelia:1997gz,AmelinoCamelia:2008qg}. 

GW propagation speed has been constrained indirectly from ultra-high energy cosmic rays. Assuming the cosmic rays originate in our Galaxy, the absence of gravitational Cherenkov radiation and the consequent observation of such cosmic rays on the Earth lead to the limit on GW speed, $c-\upsilon_g < 2 \times 10^{-15} c$ \cite{Moore:2001bv}. The constraints on anisotropic GW speed from the gravitational Cherenkov radiation have been extensively studied in the context of gravitational standard-model extension via Lorentz violation \cite{Kostelecky2015PLB}. However, the above constraint on isotropic GW speed can be applied only to subluminal case. On the other hand, from the observational data of the orbital decay of a binary pulsar, the constraint on superluminal GW speed has been obtained, $|c-\upsilon_g| \lesssim 10^{-2} c$ \cite{Jimenez2015arXiv}. Although this constaint can be applied to both super- and subluminal propagations, there is still large parameter space allowed for modification of gravity. In addition, all the constraints above are indirect measurements of the GW velocity. Therefore, the direct measurement of GW propagation speed is crucial in testing gravity theories.

So far there have been a few proposals to directly measure the GW propagation speed. One is comparing the phases of a GW and its electromagnetic counterpart from a periodic binary source \cite{Larson:1999kg,Cutler:2002ef}. However, to eliminate unknown intrinsic phase lag between the GW and the electromagnetic wave at the source, two signals at different times (e.g. a half year) on the Earth's orbit around the Sun have to be differentiated. Then the gain of the differential signal is suppressed by the propagation distance of the order of $\sim 1\,{\rm{AU}}$. A similar method using the R\o mer time delay has been suggested recently \cite{Finn:2013toa}. A GW signal from a periodic GW source is modulated in phase due to the Earth revolution. Although this method does not require any electromagnetic observation, the measurement precision is again determined basically by the baseline of the solar system. 

To extend the baseline and improve the sensitivity, in our previous work \cite{Nishizawa2014PRD}, we have reported a simple method directly comparing the arrival times between GWs, and neutrinos or photons from supernovae (SN) and short gamma-ray burst (SGRB), assuming that the SGRB is associated with a NS-NS or NS-BH binary merger \cite{Berger:2013jza}, where NS and BH represent neutron star and black hole, respectively. One might concern about unknown intrinsic time delay at the source, which depends on the emission mechanisms of GWs, neutrinos, and photons. However, numerical simulations have been well developed in these days and start to allow us to predict the intrinsic time delays. Thanks to the developments of numerical simulations, the future multimessenger observations of a GW, neutrinos, and photons can test the GW propagation speed at the precision of $\sim 10^{-15}$, improving the previous suggestions by 8-10 orders of magnitude. In this paper, we extend the previous method to a multiple-event case at cosmological distance, and show that the intrinsic time delay can be distinguished from a true signal due to anomalous GW speed by considering their redshift dependences. We also show that some combinations of signals cancel out the intrinsic time delay and give nearly optimal sensitivity. 

This paper is organized as follows. In Sec.~\ref{sec2}, we briefly review the method comparing the arrival times of a GW and a high energy photon from a SGRB in order to constrain GW propagation speed, extending the previous formalism to compact binaries at cosmological distance. In Sec.~\ref{sec3}, we introduce the framework of Bayesian inference for parameter estimation of GW propagation models. The method is numerically demonstrated in Sec.~\ref{sec4}, showing the expected constraints in the future. In Sec.~\ref{sec5}, several details of the method are discussed, taking into account more practical situations: optimality of the statistic, scaling of sensitivity, and the presence of high-$z$ cutoff for SGRB detection and its effect on sensitivity. Finally, Sec.~\ref{sec6} is devoted to a summary. In this paper, we use the unit $c=1$.

%%%%%%%%%%%%%%%%%%%%%%%%%%%%%%%%%%%%%%%%%%
\section{Arrival time delays}
\label{sec2}

Let us start with a brief review of the method comparing the arrival times of a GW and a high-energy photon from the same source to constrain GW propagation speed. As a source, in this paper we concentrate on a SGRB, assuming that the SGRB is associated with a NS-NS or NS-BH binary merger.

A GW is emitted at the time $t=t_e$ and is detected on the Earth at $t=t_e+T_g$, where the arrival time refers to, for instance, the merger time of a NS binary and $T_g$ is the propagation time of the GW from the source to the Earth. On the other hand, a $\gamma$-ray photon accompanying to the prompt emission of SGRB is emitted at $t=t_e+\tau_{\rm{int}}$ with some intrinsic time delay $\tau_{\rm int}$ and is detected at $t=t_e+\tau_{\rm{int}}+T_{\gamma}$, where $T_{\gamma}$ is the propagation time of the photon from the source to the Earth. The observable is the difference of the arrival times between the GW and the photon and is given by
\begin{equation}
\tau_{\rm{obs}} =\Delta T + \tau_{\rm{int}} \;. \label{eq1}
\end{equation}
Here we defined $\Delta T \equiv T_{\gamma} - T_g$, which vanishes when the GW propagates with the speed of light. The sign of $\Delta T$ can be both positive or negative, depending on whether the propagation speed of the GW is superluminal or subluminal, respectively.

In order that the finite time lag due to the anomalous GW speed is detectable, $\Delta T$ has to exceed uncertainties in the intrinsic time lag of the emissions, $\tau_{\rm{int,min}} \leq \tau_{\rm int} \leq \tau_{\rm{int,max}}$, and satisfy one of the following two conditions: $\tau_{\rm{int,max}} <  \Delta T + \tau_{\rm{int,min}}$ for $\Delta T >0$ and $\Delta T + \tau_{\rm{int,max}} < \tau_{\rm{int,min}}$ for $\Delta T <0$, equivalently, 
\begin{equation}
\Delta \tau_{\rm{int}} <  |\Delta T| \;,
\label{eq15a}
\end{equation}
with $\Delta \tau_{\rm{int}} \equiv \tau_{\rm{int,max}}-\tau_{\rm{int,min}}$.

Note in the derivation of Eq.~(\ref{eq15a}) that we have not taken into account the detection timing errors of a GW and a photon when they are detected on the Earth. The phase error of a GW significantly depends on the signal-to-noise ratio (SNR) and is given roughly by $\Delta \phi_{\rm{gw}} \sim {\cal{O}}({\rm{SNR}})^{-1}$ \cite{Cutler:1994ys}. For a NS binary merger detected by aLIGO, SNR is typically $\sim10$ at $200\,{\rm{Mpc}}$. Then the detection timing error of a GW is at most $\sim 10^{-3}\,{\rm{sec}}$. This is also true for ET because of a similar SNR for a NS binary even at a high redshift. Because the intrinsic uncertainty of emission time, e.g. $\sim 10\,{\rm{sec}}$ or more for SGRB photons, is much larger than the detection timing error, we can neglect it when we consider the constraint on the GW speed.

Next we derive the explicit expression of $\Delta T$, taking into account the redshift effect due to the cosmological expansion, because the third-generation ground-based GW detector such as ET enables us to observe NS-NS binaries at cosmological distances up to $z\sim2$, while for NS-BH binaries up to $z\sim 4$ \cite{Sathyaprakash:2009xt}.
Let us assume a flat Lambda Cold Dark Matter ($\Lambda$CDM) universe for simplicity. Strictly speaking, this assumption is not valid when we deal with modified gravity because dynamics of the cosmic expansion is also modified. However, to be consistent with observational data, the cosmic expansion has to be close to that in $\Lambda$CDM universe and is well approximated by $\Lambda$CDM model for our purpose here. 

The comoving distance from the observer at $z_o$ to a source at redshift $z$ is
\begin{equation}
\chi (z_o,z) = \int_{z_o}^{z} \frac{\upsilon_g}{H(z)}dz \;,
\end{equation}
and is written $\chi_0 (z_o,z)$ when the GW propagation speed is $\upsilon_g=c$. Here $H(z)$ is the Hubble parameter given by
\begin{equation}
H(z) = H_0 \sqrt{\Omega_{\rm{m}}(1+z)^3+\Omega_{\Lambda}} \;.
\end{equation}
where $\Omega_{\rm{m}}$ and $\Omega_{\Lambda}=1-\Omega_{\rm{m}}$ are the energy densities of matter and a cosmological constant, and $H_0$ is the Hubble constant at present. In this paper, we use the cosmological parameters, $H_0=100\, h_0 \,{\rm{km}}\,{\rm{Mpc}}^{-1}\,{\rm{s}}^{-1}$ with $h_0=0.68$, $h_0^2 \Omega_{\rm{m}}=0.14$ \cite{Ade:2013zuv}. It is convenient to define $\delta_g \equiv (c-\upsilon_g)/c$. The GW propagation speed $\upsilon_g$ is in general time-dependent \cite{Saltas:2014dha,Bellini2014JCAP,Gleyzes2014IJMPD} and should deviate from $c$ in the current epoch of the Universe if modification of gravity allows the GW speed to change and simultaneously explains the self-acceleration of the cosmic expansion \cite{Lombriser2015arXiv}. Motivated by these facts, we parameterize the functional form as $\delta_g=\delta_0 (1+z)^{-n}$, where $\delta_0$ is $\delta_g$ at present and $n=0$ corresponds to the constant case $\delta_g=\delta_0$. The index $n$ is different in each gravity model and has no preferred value from the observational point of view, but as pointed out in \cite{Lombriser2015arXiv} it might be increasing faster than the decrease of the matter energy density to affect the cosmic expansion of the current Universe. $n$ should not be large negative number so as not to diverge at high redshifts. Therefore, we consider in this paper the range $-1 \leq n \leq 4$. From $\chi(-\Delta z,z)=\chi_0 (0,z)$, the time delay (or advance) induced by $\delta_g$ is
\begin{equation}
\Delta T = \frac{\Delta z}{H_0} = \delta_0 \int_0^z \frac{dz}{(1+z)^{n}H(z)} \;.
\label{eq4}
\end{equation}

Also the intrinsic time delay is redshifted. Denoting the intrinsic time delay at the source as $\tilde{\tau}_{\rm{int}}$, the time delay we observe on the Earth is
\begin{equation}
\tau_{\rm{int}} (z) =(1+z)\, \tilde{\tau}_{\rm{int}} \;. 
\end{equation} 
The difference of arrival times observed on the Earth is 
\begin{equation}
\tau_{\rm{obs}} (z) = \Delta T (z) + \tau_{\rm{int}} (z)\;. 
\label{eq91}
\end{equation}

\begin{figure}[t]
\begin{center}
\includegraphics[width=8cm]{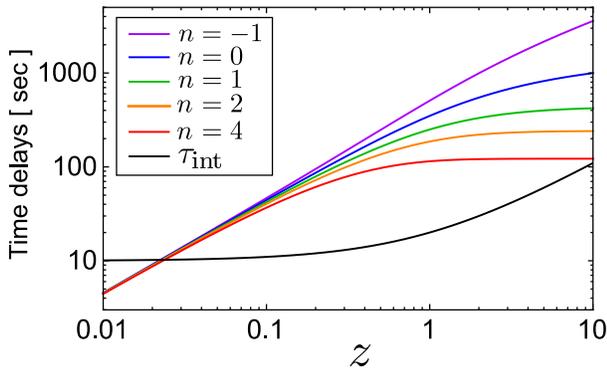}
\caption{Arrival time lags due to GW speed $\delta_g$ and the intrinsic time delay as a function of redshift. For illustration, the parameters are chosen as $\delta_g=10^{-15}$ and $\tilde{\tau}_{\rm{int}}=10\,{\rm{sec}}$.}
\label{fig4}
\end{center}
\end{figure}

In Fig.~\ref{fig4}, the GW time delay due to $\delta_g$ and intrinsic time delay are illustrated for the case of $\delta_{g}=10^{-15}$ and $\tilde{\tau}_{\rm{int}}=10\,{\rm{sec}}$. The GW time delay increases at low $z$, proportional to the distance to the source. At high $z$, however, the cosmic expansion modifies the dependence of the time delay on the distance (redshift) and the growth of the time delay slows down. As the index $n$ increases from $-1$ to $4$, the contribution of the time delay at high $z$ is more suppressed. On the other hand, the intrinsic time delay is constant at low $z$ but linearly increases at high $z$.

We define the difference of arrival times in the source frame by
\begin{equation}
\Delta \tilde{T} (z) \equiv \frac{\Delta T (z)}{1+z} \;, \quad \quad \tilde{\tau}_{\rm{obs}} (z) \equiv \frac{\tau_{\rm{obs}}(z)}{1+z} \;.
\end{equation}
Then Eq.~(\ref{eq91}) converted to in a source frame is
\begin{equation}
\tilde{\tau}_{\rm{obs}} (z) = \Delta \tilde{T} (z) + \tilde{\tau}_{\rm{int}} \;. 
\label{eq92}
\end{equation}
This expression is useful because only the signal depends on redshift, not the noise. Furthermore, for the later use, we write $\tilde{\tau}_{\rm{int}}$ as the sum of the expectation value $\langle \tilde{\tau}_{\rm{int}} \rangle$ and a fluctuating part around the expectation value $\delta \tilde{\tau}_{\rm{int}}$. Then Eq.~(\ref{eq92}) can be separated into the systematic and statistical terms:
\begin{align}
\tilde{\tau}_{\rm{obs}} (z) &= \langle \tilde{\tau}_{\rm{obs}} (z) \rangle + \delta \tilde{\tau}_{\rm{obs}} \;, 
\label{eq14} \\
\langle \tilde{\tau}_{\rm{obs}} (z) \rangle &\equiv \Delta \tilde{T} (z) + \langle \tilde{\tau}_{\rm{int}} \rangle \;, 
\label{eq93} \\
\delta \tilde{\tau}_{\rm{obs}} &\equiv \delta \tilde{\tau}_{\rm{int}}\;. 
\end{align}

%%%%%%%%%%%%%%%%%%%%%%%%%%%%%%%%%%%%%%%%%%
\section{Bayesian inference}
\label{sec3}

The detections of multiple events at cosmological distance would be realized with the third-generation ground-based GW detector such as ET. From the consideration of the beaming angle of SGRB \cite{Nishizawa:2011eq}, more than several tens of GW-SGRB coincidence events would be observed with ET and gamma-ray detectors in a realistic observation time, e.g. 1\,{\rm{yr}}. With these coincidence events, one can distinguish the true signal due to finite $\delta_g$ from the intrinsic time delay of the emission at a source by utilizing their redshift dependences. To utilize multiple coincidence events of NS-NS binaries or NS-BH binaries and SGRB for measuring the propagation speed of a GW, we introduce the framework of Bayesian inference to estimate errors in model parameters of GW propagation.

According to the Bayes theorem, the posterior probability distribution is given by
\begin{equation}
p(\vec{\theta} | D, {\cal{H}}) = \frac{p(D|\vec{\theta},{\cal{H}}) p(\vec{\theta}|{\cal{H}})}{p(D|{\cal{H}})} \;,
\label{eq:Bayes}
\end{equation}
where $\vec{\theta}$ is a set of model parameters, ${\cal{H}}$ is a hypothesis, and $D$ is observational data. On the right-hand side of Eq.~(\ref{eq:Bayes}), $p(D|\vec{\theta},{\cal{H}})$ is the likelihood, $p(\vec{\theta}|{\cal{H}}))$ is the prior distribution, and $p(D|{\cal{H}})$ is the evidence. The evidence is merely a normalization factor of the posterior probability distribution and does not affect physical consequences. 

We assume that the statistical fluctuation of the intrinsic time delay obeys the Gaussian distribution whose variance is given by $\sigma_{\tau}^2 = \left\langle \left (\delta \tilde{\tau}_{\rm{int}} \right)^2 \right\rangle$. This assumption is equivalent to writing the unnormalized likelihood probability of a single event using Eq.~(\ref{eq14}) as
\begin{equation}
\exp \left[ -\frac{\left\{ \delta \tilde{\tau}_{{\rm obs},i} \right\}^2}{2\sigma_{\tau}^2} \right] = \exp \left[ -\frac{\left\{ \tilde{\tau}_{\rm{obs}} (z_i) - \langle \tilde{\tau}_{\rm{obs}} (z_i) \rangle \right\}^2}{2\sigma_{\tau}^2} \right] \;,
\end{equation}
where the index $i$ discriminates each event. Since each event is independent one another, the total likelihood is 
\begin{align}
p(D|\vec{\theta},{\cal{H}}) &\propto \prod_i \exp \left[ -\frac{\left\{ \tilde{\tau}_{\rm{obs}} (z_i) - \langle \tilde{\tau}_{\rm{obs}} (z_i) \rangle \right\}^2}{2\sigma_{\tau}^2} \right] \nonumber \\
&= \exp \left[ - \sum_i \frac{\left\{ \tilde{\tau}_{\rm{obs}} (z_i) - \langle \tilde{\tau}_{\rm{obs}} (z_i) \rangle \right\}^2}{2\sigma_{\tau}^2} \right]  \;.
\label{eq:likelihood}
\end{align}

In our case, the hypothesis ${\cal{H}}$ is that the Universe is described by flat $\Lambda$CDM model. However, the cosmological parameters in the flat $\Lambda$CDM model, $H_0$ and $\Omega_{\rm m}$, are well determined within $5\%$ precision from the cosmological observations \cite{Ade:2013zuv} and their uncertainties do not much affect the errors in the measurement of GW propagation speed. Thus, we exclude $H_0$ and $\Omega_{\rm m}$ from free parameters in our analysis and take $\vec{\theta}=\{ \delta_0, n, \langle \tilde{\tau}_{\rm int} \rangle \}$ as free parameters. In other words, the priors on $H_0$ and $\Omega_{\rm m}$ are regarded as the delta functions. On the other hand, we apply flat priors for 
$\delta_0$, $n$, and $\langle \tilde{\tau}_{\rm int} \rangle$. Our fiducial values for the model parameters are $\delta_0=0$, $n=0$, $\langle \tilde{\tau}_{\rm int} \rangle=150\,{\rm sec}$. The choice of $\langle \tilde{\tau}_{\rm int} \rangle=150\,{\rm sec}$ might seem to be intentional. However, as discussed in Sec.~\ref{sec5a}, it is irrelevant to constrain the GW speed because it can always be canceled out by pairing the signals.   

The magnitude of a measurement noise in the time-delay signal is determined by $\sigma_{\tau}$, which depends on the emission mechanism of SGRB. In this paper, we consider three cases: $\sigma_{\tau}=10, 25, 50\,{\rm sec}$. The reason of these choice is because the duration of SGRB is typically less than $\sim 2\,{\rm sec}$ and the fluctuations of $\tilde{\tau}_{\rm int}$ is expected to be the same order of magnitude or less from consideration of the emission mechanisms \cite{LiHu2016arXiv}. However, to be conservative, we consider not only $10\,{\rm sec}$ but also larger noises $25\,{\rm sec}$ and $50\,{\rm sec}$. 

When $\delta_g$ is a time-varying function and contains two free parameters, it is convenient to show the posterior distribution by marginalizing over $\langle \tilde{\tau}_{\rm int} \rangle$. The marginalized distribution can be derived as follows. We write $q \equiv \langle \tilde{\tau}_{\rm int} \rangle$ and $\hat{q}_i \equiv \tilde{\tau}_{\rm{obs}} (z_i) - \Delta \tilde{T} (z_i)$ for simplicity of notation. From Eqs.~(\ref{eq93}), (\ref{eq:Bayes}), and (\ref{eq:likelihood}), the marginalized posterior distribution is 
\begin{align}
p(\vec{\theta}^{\prime} | D, {\cal{H}}) &\propto \int dq \exp \left[ - \sum_i \frac{\left\{ \hat{q}_i - q \right\}^2}{2\sigma_{\tau}^2} \right] \nonumber \\
&= \exp \left[ -\frac{1}{2\sigma_{\tau}^2} \left( \hat{Q}_2 -\frac{\hat{Q}_1^2}{N_{\rm total}} \right)  \right] \nonumber \\
&\times \int dq \exp \left[  -\frac{1}{2\sigma_{\tau}^2} N_{\rm total} \left( q-\frac{\hat{Q}_1}{N_{\rm total}} \right)^2 \right] \nonumber \\
&\propto \exp \left[ -\frac{1}{2\sigma_{\tau}^2} \left( \hat{Q}_2 -\frac{\hat{Q}_1^2}{N_{\rm total}} \right) \right] \;,
\label{eq94} 
\end{align}
\begin{equation}
\hat{Q}_1 \equiv \sum_i \hat{q}_i \;, \quad \quad \hat{Q}_2 \equiv \sum_i \hat{q}_i^2 \;, \nonumber 
\end{equation}
where $\vec{\theta}^{\prime}=\delta_0, n$ and $N_{\rm total}$ is the total number of sources. Particularly, when $N_{\rm total} \rightarrow \infty$, $\hat{Q}_1/N_{\rm total}$ approaches the expectation value $\bar{q}$. Therefore,
\begin{align}
p(\vec{\theta}^{\prime} | D, {\cal{H}}) &\propto \exp \left[ -\frac{1}{2\sigma_{\tau}^2} \sum_i \hat{q}_i \left( \hat{q}_i- \bar{q} \right) \right] \nonumber \\
&= \exp \left[ -\frac{1}{2\sigma_{\tau}^2} \left\{ \sum_i \left( \hat{q}_i- \bar{q} \right)^2 \right. \right. \nonumber \\
& \left. \left. \quad \quad \quad +\bar{q} \left( \hat{Q}_1 - N_{\rm total}\, \bar{q} \right) \right\} \right] \;. \label{eq:mlikelihood}
\end{align}
By the definition of the expectation value, the second term in the bracket vanishes. Thus, the marginalized posterior distribution obeys the Gaussian distribution with respect to $\hat{q}_i$. Namely, the logarithmic posterior distribution marginalized over $\langle \tilde{\tau}_{\rm int} \rangle$ obeys $\chi^2$ distribution. The above result is derived for infinite $N_{\rm total}$. However, it is expected that Eq.~(\ref{eq:mlikelihood}) also holds for the large number of sources.

%%%%%%%%%%%%%%%%%%%%%%%%%%%%%%%%%%%%%%%%
\section{Numerical Implementation}
\label{sec4}

In this section, we numerically generate mock data of events and investigate expected constraints on model parameters, $\delta_0$, $n$, and $\langle \tilde{\tau}_{\rm int} \rangle$, based on the Bayesian approach.

%%%%%%%%%%%%%%%%%%%%%%%%%%%%%%%%%%%%%%%%%%
\subsection{Procedures}
\label{sec4a}
The procedures of data analysis are composed of three stages.

\begin{enumerate}
\item{Redshift distribution of NS binary merger events}

\begin{figure}[t]
\begin{center}
\includegraphics[width=8cm]{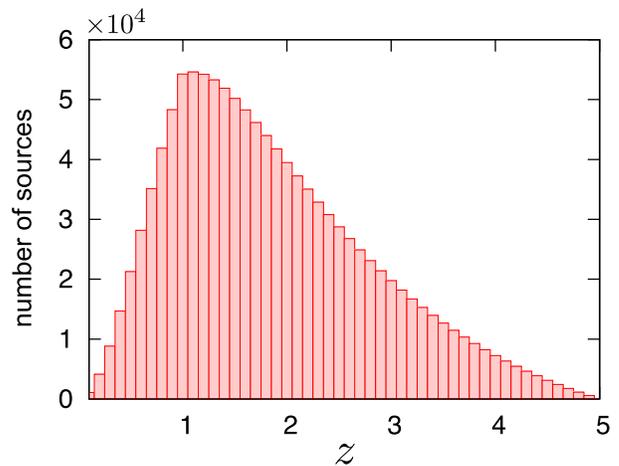}
\caption{Number of NS-NS binaries (in the unit of $10^4$) in each redshift bin of $\Delta z =0.1$ at a redshift $z$ during $1\,{\rm{yr}}$ observation.}
\label{fig6}
\end{center}
\end{figure}

$\dot{n}(z)$ is the NS merger rate per unit comoving volume per unit proper time at a redshift $z$. The fitting formula based on the observation of star formation history is given in \cite{Cutler:2006} by  
\begin{equation}
\dot{n}(z) = \dot{n}_0 \times \left\{        
\begin{array}{ll}
1+2 z  & (z \leq 1) \\
\frac{3}{4} (5-z) &(1 < z \leq 5) \\
 0 & (5 < z)
\end{array}
\right. \;, 
\label{eq:ns-dist-rate}
\end{equation}
where the quantity $\dot{n}_0$ represents the merger rate at present. Although the normalization of $\dot{n}$ is still largely uncertain, we adopt the intermediate value of recent estimates, $\dot{n}_0=10^{-6}\,{\rm{Mpc}}^{-3}\, {\rm{yr}}^{-1}$, as a reliable estimate based on extrapolations from the observed binary pulsars in our Galaxy \cite{Abadie:2010cf}. The number of NS binary merger in the redshift interval $[z,z+dz]$ observed during the observation time $T_{\rm obs}$ is given by \cite{Cutler:2006}
\begin{equation}
\frac{dN(z)}{dz}=T_{\rm obs}\,\frac{4\pi r^2(z)}{H(z)}\frac{\dot{n}(z)}{1+z} \;, 
\label{eq:ns-dist-z}
\end{equation}
where $r(z)$ is the comoving radial distance and is related to the luminosity distance $d_L(z)$ by $r(z)=d_L(z)/(1+z)$ in the flat universe. In Fig.~\ref{fig6}, using Eq.~(\ref{eq:ns-dist-z}), the redshift distribution of NS binaries per year is shown.

To generate NS binary merger events that obeys the redshift distribution in Eq.~(\ref{eq:ns-dist-z}) from a homogeneous random distribution, we use the Box-Muller method \cite{Press1992book}. 
%By linearly fitting the distribution in Fig.~\ref{fig6}, the cumulative number of NS binary mergers up to the redshift $z$ is approximately given by
%\begin{align}
%N(z) &= \int_{0}^{z} dz^{\prime} \frac{dN(z^{\prime})}{dz^{\prime}} \nonumber \\
%&= N_0 \left( T_{\rm obs}/ 1\,{\rm yr} \right)  \nonumber \\
%& \times \left\{        
%\begin{array}{ll}
%z^2  & (z \leq 1) \\
%-1.4+2.7z-0.34 z^2 &(1 < z \leq 3)
%\end{array}
%\right. \;, \nonumber \\
%\end{align}
%where $N_0=2.3 \times 10^4$. 
In our numerical simulation, we take into account NS binary merger events only at the redshift range $z<2$, because the electromagnetic identification of SGRB at higher redshifts would be difficult and it seems to be realistic to assume that sources at $z<2$ can be identified as coincident events between electromagnetic waves and GWs. We denote the fraction of coincidence events among all NS binary merger events by $\epsilon$ and use $\epsilon=10^{-3}$, which is estimated from the simple consideration of SGRB jet opening angle \cite{Nishizawa:2011eq}. Thus, the cumulative number of coincidence events out to a redshift $z$ is $\epsilon N(z)$ and the total number of coincidence events is $N_{\rm total}=\epsilon N(z_{\rm max})$, where $N(z)$ is the cumulative number of GW events out to a redshift $z$ and $z_{\rm max}$ is the maximum redshift that an electromagnetic counterpart of a GW source is detected.

\item{Generating time delay signals}

Time delay signals are generated using Eq.~(\ref{eq91}) for fixed parameters $\delta_0$, $n$, and $\langle \tilde{\tau}_{\rm int} \rangle$. The error of the intrinsic time delay is added to each signal by generating a Gaussian error with the standard deviation $\sigma_{\tau}$, for which we choose $\sigma_{\tau}=10, 25, 50\,{\rm sec}$. We fix the expectation value of an intrinsic time delay to $\left\langle \tilde{\tau}_{\rm{int}} \right\rangle=150\,{\rm sec}$. However, this does not loose generality because as discussed in Sec.~\ref{sec5a} the expectation value of an intrinsic time delay can be canceled by taking the difference of two signals at different redshifts.

\item{Computation of the posterior distribution}

Since we apply flat priors for $\delta_0$, $n$, and $\langle \tilde{\tau}_{\rm int} \rangle$, the posterior distribution is obtained from the likelihood distribution in Eq.~(\ref{eq:likelihood}) except for its normalization. The posterior distribution marginalized over $\langle \tilde{\tau}_{\rm int} \rangle$ is given by Eq.~(\ref{eq94}). From these posterior distributions, we compute parameter estimation errors at 68\% CL. To suppress a sampling error, we average the parameter estimation errors over $100$ realizations of the event list. As a result, the averaged constraints are less fluctuating, but still fluctuate by $\sim 5\%$, at most $10\%$. 

\end{enumerate}

%%%%%%%%%%%%%%%%%%%%%%%%%%%%%%%%%%%%%%%%%%
\subsection{Expected errors of model parameters}

\begin{figure}[t]
\begin{center}
\includegraphics[width=8cm]{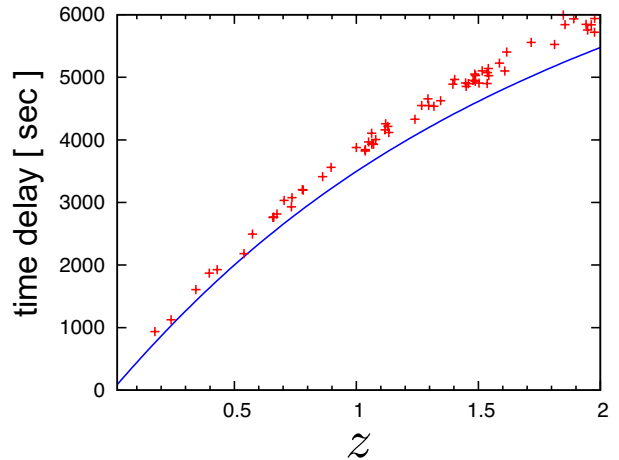}
\caption{One realization of time delay signals as a function of redshift when $\delta_g=10^{-14}$ ($n=0$), $\sigma_{\tau}=50\,{\rm{sec}}$, and $\epsilon=10^{-3}$. The red points are mock time-delay signals and the dashed curve is the time delay due to finite $\delta_g$.}
\label{fig8}
\end{center}
\end{figure}

In Fig.~\ref{fig8}, the generated time-delay signals of events are plotted as a function of redshift. Just for the illustrative purpose, the parameters are chosen as $\delta_g=10^{-14}$, $\sigma_{\tau}=50\,{\rm{sec}}$, and $\epsilon=10^{-3}$ ($N_{\rm total}=63$). It is seen that the intrinsic time delay in the observer's frame is redshifted and larger at high $z$ and that the more sources are distributed at redshifts from 1 to 1.5 as expected from the redshift distribution in Fig.~\ref{fig6}.

\begin{figure}[t]
\begin{center}
\includegraphics[width=6.5cm]{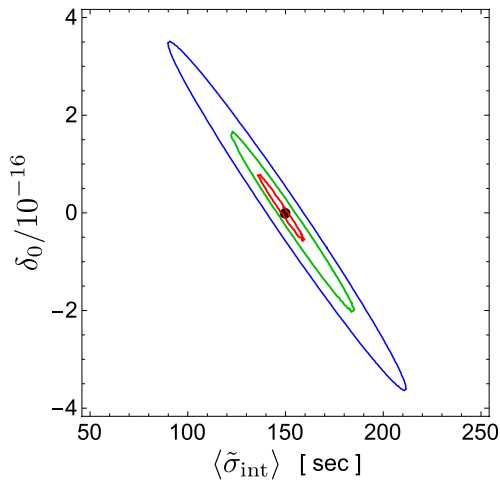}
\caption{Constraint on constant case $\delta_g=\delta_0$ and $\langle \tilde{\tau}_{\rm int} \rangle$ when $\epsilon=10^{-3}$ ($N_{\rm total}=63$). The fiducial parameters are chosen $\delta_g=\delta_0=0$ and $\langle \tilde{\tau}_{\rm int} \rangle=150\,{\rm sec}$, represented by a black point at the center of the figure. From the smaller ellipses to the larger, the fluctuations of intrinsic time delays are $\sigma_{\tau}=10$, $25$, and $50\,{\rm sec}$.}
\label{fig12}
\end{center}
\end{figure}

The posterior distribution of $\delta_0$ and $\langle \tilde{\tau}_{\rm int} \rangle$ in the case of constant $\delta_g$ ($n=0$) is shown in Fig.~\ref{fig12}. The errors in $\delta_0$ and $\langle \tilde{\tau}_{\rm int} \rangle$ are strongly correlated. This is because the larger $\langle \tilde{\tau}_{\rm int} \rangle$ is equal to negative $\delta_0$ (superluminal propagation) in the observational signal in Eq.~(\ref{eq91}). However, they do not completely degenerate because of different redshift dependence. The expected constraints on $\delta_g$ (68\% CL) are $-0.6< \delta_g/10^{-16} <0.8$, $-2.0< \delta_g/10^{-16} <1.7$, and $-3.6< \delta_g/10^{-16} <3.5$ for $\sigma_{\tau}=10$, $25$, and $50\,{\rm sec}$.

\begin{figure}[h]
\begin{center}
\includegraphics[width=8.5cm]{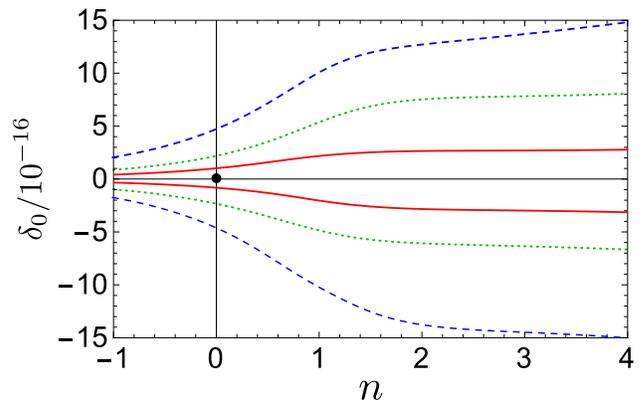}
\caption{Constraint on $\delta_0$ and $n$ when $\langle \tilde{\tau}_{\rm int} \rangle$ distribution is marginalized and $\epsilon=10^{-3}$ ($N_{\rm total}=63$). The fiducial parameters are chosen $\delta_0=0$, $n=0$, and $\langle \tilde{\tau}_{\rm int} \rangle=150\,{\rm sec}$, represented by a black point at the center of the figure. The fluctuations of intrinsic time delays are $\sigma_{\tau}=10$ (red, solid), $25$ (green, dotted), and $50\,{\rm sec}$ (blue, dashed).}
\label{fig13}
\end{center}
\end{figure}

In Fig.~\ref{fig13}, the posterior distribution marginalized over $\langle \tilde{\tau}_{\rm int} \rangle$ is shown. The constraint is tighter at smaller $n$ just because of the redshift dependence of $\delta_g$. When $n$ is negative, the absolute value of $\delta_g$ increases at higher redshifts. On the other hand, when $n$ is positive, $|\delta_g|$ is suppressed at higher redshifts and becomes more difficult to detect. In Table~\ref{tab1}, the projected constraints on $\delta_0$ for different $\sigma_{\tau}$ and $n$ are listed. It should be noted that the constraints on $\delta_0$ for $n\neq 0$ are those obtained when the fiducial parameters are $\delta_0=0$ and $n=0$. In other words, those are what is derived from the data when no positive detection is achieved and true parameters are $\delta_0=0$ and $n=0$.

\begin{table}[t]
\begin{center}
\begin{tabular}{|c|c|c|c|}
\hline
& $\sigma_{\tau}=10\,{\rm sec}$ & $\sigma_{\tau}=25\,{\rm sec}$ & $\sigma_{\tau}=50\,{\rm sec}$ \\
\hline 
$n=-1$ & $-0.3 < \delta_0 < 0.4$ & $-1.0 < \delta_0 < 0.9$ & $-1.8 < \delta_0 < 2.0$ \\
\hline 
$n=0$ & $-0.8 < \delta_0 < 1.0$ & $-2.2 < \delta_0 < 2.0$ & $-4.8 < \delta_0 < 4.7$ \\
\hline 
$n=1$ & $-2.1 < \delta_0 < 2.1$ & $-4.9 < \delta_0 < 5.3$ & $-10.2 < \delta_0 < 10.0$ \\
\hline
$n=2$ & $-2.9 < \delta_0 < 2.6$ & $-6.1 < \delta_0 < 7.5$ & $-13.8 < \delta_0 < 12.8$ \\
\hline 
$n=4$ & $-3.1 < \delta_0 < 2.9$ & $-6.8 < \delta_0 < 8.0$ & $-15.0 < \delta_0 < 14.9$ \\
\hline 
\end{tabular}
\end{center}
\caption{Expected constraint on $\delta_0$ (68\% CL) for different $\sigma_{\tau}$ and $n$ in the redshift-dependent $\delta_g$ case with fiducial parameters $\delta_0=0$ and $n=0$. The values of $\delta_0$ is in the unit of $10^{-16}$.}
\label{tab1}
\end{table}

%%%%%%%%%%%%%%%%%%%%%%%%%%%%%%%%%%%
\section{Discussions}
\label{sec5}

In this section, we focus on the case of constant $\delta_g$ ($n=0$) and investigate physical aspects of sensitivity and a concrete statistic to interpret the results.

%%%%%%%%%%%%%%%%%%%%%%%%%%%%%%%%%%%%%%%%%%
\subsection{Optimal statistic}
\label{sec5a}

\begin{figure}[t]
\begin{center}
\includegraphics[width=8cm]{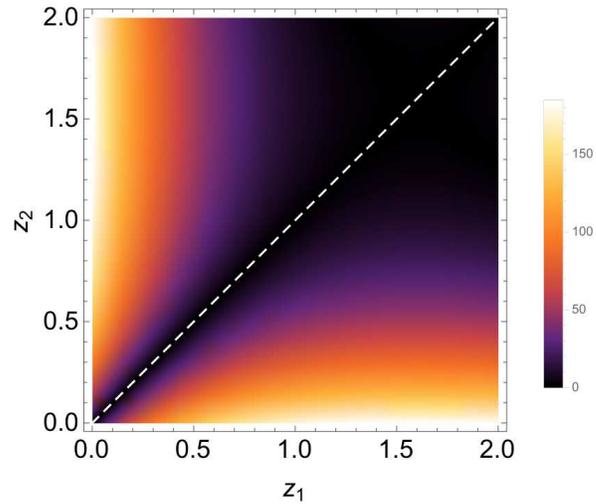}
\caption{A differential signal of arrival time delays $| \langle s(z_1,z_2) \rangle |$ in the unit of sec when $\delta_g=10^{-15}$ ($n=0$). The diagonal line is $z_1=z_2$.}
\label{fig1}
\end{center}
\end{figure}

\begin{figure}[t]
\begin{center}
\includegraphics[width=8cm]{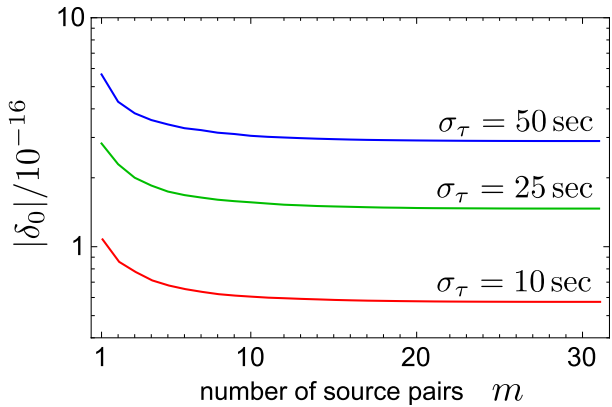}
\caption{Constraints to constant $\delta_g=\delta_0$ as a function of the number of event pairs for $\sigma_{\tau}=50$, $25$, and $10\,{\rm sec}$ from the top to the bottom, respectively. $m$ represents event pairs and runs from the largest redshift-separation pair to the smallest one.}
\label{fig9}
\end{center}
\end{figure}

If one has multiple SGRB events observed coincidentally by GW and $\gamma$-ray detectors, one can distinguish the true signal due to finite $\delta_g$ and the intrinsic time delay at a source by looking at the redshift dependence. To do so, we consider a new statistic that could be used in a real data analysis. The observed quantity is the arrival time delay $\tilde{\tau}_{\rm{obs}}$, from which we can construct the following statistic:
\begin{align}
s (z_i,z_j) &\equiv \tilde{\tau}_{\rm{obs}} (z_i) - \tilde{\tau}_{\rm{obs}} (z_j) \nonumber \\
&= \Delta \tilde{T}(z_i)-\Delta \tilde{T}(z_j) + \delta \tilde{\tau}_{{\rm{int}},i}- \delta \tilde{\tau}_{{\rm{int}},j} \;. \label{eq7}
\end{align}
where $i$ and $j$ denote $i$-th and $j$-th events. The second term is stochastic with zero mean, while there remains finite contribution from GW. Therefore, we have 
\begin{equation}
\left\langle s (z_i,z_j) \right\rangle = \Delta \tilde{T}(z_i)-\Delta \tilde{T}(z_j) \;. 
\label{eq7a}
\end{equation}
\begin{align}
{\rm Var} [ s (z_i,z_j)] &= \langle (s (z_i,z_j)-\langle s (z_i,z_j) \rangle)^2 \rangle \nonumber \\
&= 2 \langle \delta \tilde{\tau}_{{\rm{int}}}^2 \rangle \nonumber \\
&= 2 \sigma_{\tau}^2
\end{align}
In Fig.~\ref{fig1}, we show the redshift dependence of $| \langle s (z_i,z_j) \rangle|$. Since the noise $\delta \tilde{\tau}_{{\rm{int}}}$ does not depend on a redshift, the redshift dependence of the SNR is identical to that of a signal. This implies two crucial facts to construct an optimal statistic. Firstly, it hardly depends on the redshift for $z>1$. In other words, high-$z$ sources at $z \gtrsim 2$, for which it is more difficult to have an electromagnetic counterpart, do not play an important role in obtaining large SNR. Secondly, since $\Delta \tilde{T} (z)$ is monotonously increasing (decreasing) function for positive (negative) $\delta_g$ below $z=1$, the largest SNR is obtained by taking the difference of time delays at largely separating redshifts, $|z_i-z_j| \gtrsim 1$. Thus, when one has multiple events, the tightest constraint on $\delta_g$ would be imposed by a part of event pairs whose redshift difference is large.

One possible way to combine all signals at different redshifts is summing the signals over $z_i>z_j$. However, this is suboptimal because the signals are redundantly added. Indeed, for the event pairs with small redshift separation, the signals are canceled out and only noises are added. Then SNR is not improved at all. Thus, the efficient way of the summation is pairing the events from the highest and lowest redshifts and adding them in turn. This order of summation is also computationally efficient to reach the maximum sensitivity on $\delta_g$ and would be useful in a practical data analysis. The SNR for all pair of events is 
\begin{equation}
{\rm{SNR}}^2 = \sum_m \left[ \frac{\sum \left\langle s_m \right\rangle}{\sqrt{2} \, \sigma_{\tau}} \right]^2 \;, 
\label{eq95}
\end{equation}
where $m$ represents event pairs and runs from the largest redshift-separation pair to the smallest one.

We numerically generate mock data of events the same way as in Sec.~\ref{sec4a} to show explicitly that the new statistic is efficient in computation and gives almost optimal constraint on $\delta_g$ with the small number of signal pairs. In Fig.~\ref{fig9}, constraints on $\delta_g=\delta_0$ ($n=0$) as a function of the number of event pairs for different $\sigma_{\tau}$ are shown. It indicates that the SNR is dominated by only several event pairs $m$ with large redshift separation and is not improved by adding event pairs with smaller redshift separation. These asymptotic values of the constraints on $\delta_g=\delta_0$ agree well with the error ellipses in Fig.~\ref{fig12}. This means that the statistic introduced in this subsection has almost optimal sensitivity to $\delta_g$.

%%%%%%%%%%%%%%%%%%%%%%%%%%%%%%%%%%%%%%%%%%
\subsection{Scaling of SNR}

From some consideration about signal and noise, we can derive scaling relations with model parameters. Since the observable is given by Eq.~(\ref{eq7}), we do not have to care about $\langle \tilde{\tau}_{{\rm{int}}} \rangle$. Only noise scales with $\sigma_{\tau}$ and the SNR scales with $\sigma_{\tau}^{-1}$ from Eq.~(\ref{eq95}). Then the constraint on $\delta_g$ linearly scales with $\sigma_{\tau}$. As for the number of events or $\epsilon$, the scaling of $\delta_g$ is simply $\epsilon^{-1/2}$ because $\epsilon$ does not change the redshift dependence of the source distribution but its normalization. Therefore, the scaling relation for the constraint on $\delta_0$ in the case of constant $\delta_g$ is
\begin{equation}
|\delta_0| \leq 6\times 10^{-17} \left( \frac{10^{-3}}{\epsilon} \right)^{1/2} \left( \frac{\sigma_{\tau}}{10\,{\rm{sec}}} \right) \;.
\end{equation}
This formula agrees well with the errors in Fig.~\ref{fig9} and the errors from Bayesian inference in Fig.~\ref{fig12} except for some statistical fluctuations. The scaling also holds for the case of nonzero $n$. In Table~\ref{tab1}, the scaling of the constraints with $\sigma_{\tau}$ agree well. However, the magnitudes of the errors deteriorate because of some parameter degeneracies.

%%%%%%%%%%%%%%%%%%%%%%%%%%%%%%%%%%%%%%%%%%
\subsection{Maximum redshift dependence}

\begin{figure}[t]
\begin{center}
\includegraphics[width=8cm]{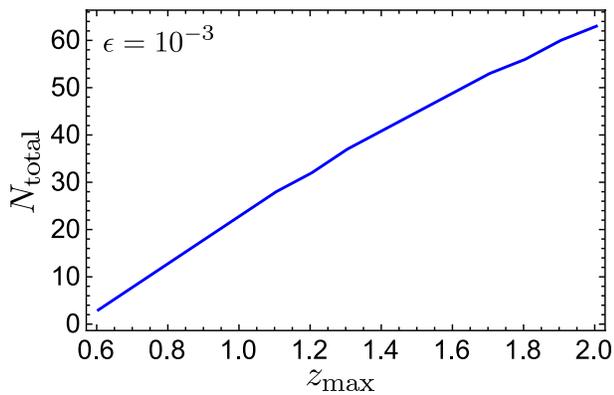}
\caption{The total number of sources up to $z=z_{\rm max}$ when $\epsilon=10^{-3}$.}
\label{fig11}
\end{center}
\end{figure}

\begin{figure}[t]
\begin{center}
\includegraphics[width=8cm]{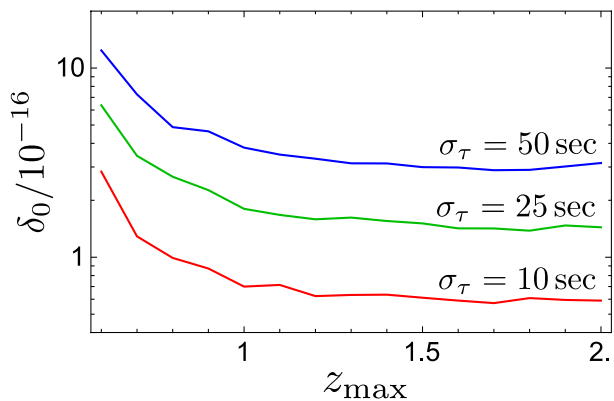}
\caption{Constraint on constant $\delta_g=\delta_0$ as a function of $z_{\rm{max}}$ for $\sigma_{\tau}=50$, $25$, and $10\,{\rm sec}$ from the top to the bottom, respectively. $\epsilon=10^{-3}$.}
\label{fig10}
\end{center}
\end{figure}

It may happen that SGRB events are seen only at low redshifts, having low-$z$ cutoff at $z<2$. In this case, the constraint on $\delta_g$ is degraded in two ways. Firstly, the number of sources decreases, as shown in Fig.~\ref{fig11} as a function of maximum redshift $z_{\rm max}$. Secondly, the signal $\left\langle s (z_i,z_j) \right\rangle$ in Eq.~(\ref{eq7a}) is likely to be small due to lack of high-$z$ sources. By these two effects, the constraint is degraded in a nontrivial way as $z_{\rm max}$ decreases. As shown in the Fig.~\ref{fig10}, the sensitivity to $\delta_g$ is drastically degraded if there is a cutoff at the redshift less than $z=1$. However, interestingly, the degradation is modest for the cutoff at $z>1$ because SNR is almost constant for sources at $z>1$, as shown in Fig.~\ref{fig1}. Therefore, we conclude that we do not necessarily have to see high-$z$ SGRBs around $z=2$ or higher, but those at $1 \lesssim z \lesssim 1.5$ are crucial.  

%%%%%%%%%%%%%%%%%%%%%%%%%%%%%%%%%%%%
\section{Conclusion}
\label{sec6}

In this paper, we have extensively studied the method measuring the GW propagation speed by directly comparing the arrival times between GWs and photons from NS binary mergers associated with SGRB. Particularly we have considered multiple coincidence events at cosmological distance, the redshift distribution of GW sources, redshift-dependent GW propagation speed, and the statistics of intrinsic time delays. Based on the Bayesian parameter inference in the realistic observational situation with ET, we have obtained the expected constraints on $\delta_g$ (68\% CL): $-0.6< \delta_g/10^{-16} <0.8$, $-2.0< \delta_g/10^{-16} <1.7$, and $-3.6< \delta_g/10^{-16} <3.5$ when $\sigma_{\tau}=10$, $25$, and $50\,{\rm sec}$, respectively, for constant $\delta_g$ ($n=0$), and the similar values of the same order in Table~\ref{tab1} for time-varying GW propagation speed (nonzero $n$). Furthermore, we have proposed an optimal statistic that  would be useful in a real data analysis. From numerical investigation of this statistic, we have shown that a systematic part of the intrinsic time delay can be canceled out from signals, distinguishing it from a true signal due to finite $\delta_g$, and that the statistic gives nearly optimal sensitivity. We also have shown that by changing the maximum redshift below which coincidence events are available, high-$z$ SGRB around $z=2$ or higher affect the sensitivity modestly, while those at $1 \lesssim z \lesssim 1.5$ are crucial in constraining $\delta_g$. 

Finally we comment on constraint on GW propagation at much higher redshifts. As a measurement method of GW propagation speed other than the one using difference of arrival times, there is a suggestion that GW speed different from $c$ at high redshifts, $\sim 10^3$, affects the cosmic microwave background (CMB) spectrum and can be measured indirectly \cite{Raveri:2014eea,Amendola:2014wma}. Since the detection of B-mode polarization by BICEP 2 turned out to be caused by dust emissions, GW speed has not been measured by the CMB observation. However, since the method with CMB is complementary to the method with arrival times from the view of redshift, both methods would give a tight constraint on the redshift evolution of the GW propagation speed in the future observations. 

%\appendix 

%%%%%%%%%%%%%%%%%%%%%%%%%%%%%%%%%%%%%%%%%%
\begin{acknowledgments}
The author would like to thank G. Ballesteros, Y. Fan, J. B. Jimenez, E. Malec, and J. D. Tasson for valuable comments. A. N. was supported by JSPS Postdoctoral Fellowships for Research Abroad.
\end{acknowledgments}

\bibliography{/Volumes/USB-MEMORY/my-research/bibliography}

\end{document}